\documentclass[a4paper]{jpconf}
\usepackage{graphicx}
\usepackage{amsfonts,amssymb}
\usepackage{amsmath}
\usepackage{hyperref,slashed,color,graphicx,url}
\hypersetup{colorlinks,citecolor= blue,linkcolor= blue, urlcolor=blue}
\begin{document}
\title{Sterile neutrino dark matter,  neutrino secret self-interactions and extra radiation}

\author{Manibrata Sen}

\address{Max-Planck-Institut f\"ur Kernphysik,
Saupfercheckweg 1, 69117 Heidelberg, Germany}

\ead{manibrata@mpi-hd.mpg.de}

\begin{abstract}
keV scale sterile neutrinos are excellent warm dark matter candidates. In the early Universe, these can be produced from oscillations and scatterings of active neutrinos. However, in the absence of any new physics, this mechanism is in severe tension with observations from X-ray searches. In this work, we show that secret self-interactions of active neutrinos, mediated by a scalar, can efficiently produce these sterile neutrinos without being ruled out by X-ray observations. These neutrino self-interactions are also testable: for a mediator mass greater than a few MeV, these self-interactions can give signatures in laboratory based experiments, while for lighter mediators, there will be observable consequences for upcoming cosmology probes. \footnote{ \footnotesize This work is based on a talk given at TAUP 2021 and a video of it can be found at this \href{https://youtu.be/Lqs_cXdWKCo}{link}.}
\end{abstract}

\section{Introduction}
The presence of non-zero neutrino masses, confirmed by neutrino oscillation data, presents robust evidence of the existence of physics beyond the Standard Model (SM) of particle physics. The SM predicts massless neutrinos, hence extensions are required to incorporate massive neutrinos. An elegant solution is to add a gauge-singlet fermion to the SM, termed as the sterile neutrino. While the explanation of neutrino masses generally require heavier fermions, lighter mass sterile neutrinos have been invoked to solve tensions arising in short baseline neutrino experiments, reactor neutrino experiments, as well as cosmological observations.

In particular, a keV-scale sterile neutrino is an excellent warm dark matter candidate (WDM). Typically characterized by a tiny mixing with the active neutrinos, these WDM can be produced non-thermally due to active-sterile oscillations in the early Universe, as was shown by Dodelson and Widrow (DW)~\cite{Dodelson:1993je}. However, the same mixing allows these sterile neutrinos to decay radiatively, producing an X-ray line~\cite{Pal:1981rm}. Non-observation of such X-ray lines, or a diffuse X-ray background near galaxy clusters leads to this model being in tension with current astrophysical bounds.

One simple way out of this impasse is to predict a large lepton asymmetry in the early Universe, which leads to a resonant production of these sterile neutrinos~\cite{Shi:1998km}. However, presence of a large lepton asymmetry is difficult to test, and, in general, difficult to explain, given the tiny baryon asymmetry. To this end, we propose a new experimentally testable way of producing the correct relic density in the Universe, without getting into trouble with astrophysical bounds~\cite{DeGouvea:2019wpf,Kelly:2020pcy,Kelly:2020aks}. We postulate ``secret" neutrino self-interactions, mediated by a scalar mediator, which can assist in efficiently producing sterile neutrino dark matter in the early Universe using the DW mechanism. These neutrino self-interactions are also testable: for a mediator mass greater than a few MeV, these self-interactions can give signatures in laboratory based experiments, while for lighter mediators, this can modify the standard cosmology, and have observable consequences for upcoming cosmology probes. 

\section{The Dodelson-Widrow mechanism}
We consider a fourth neutrino mass eigenstate $\nu_4$, which is a linear combination of the sterile state $(\nu_s)$ and the active state $(\nu_a)$, such that $\nu_4 = \cos\theta\, \nu_s + \sin\theta \, \nu_a \,$ where $\theta\ll 1$ is the active-sterile mixing angle. To begin with, the $\nu_s$ population is negligible, while the $\nu_a$ is in thermal equilibrium with the plasma. As the Universe evolves, the $\nu_a$ undergo weak-interactions with the leptons, while the $\nu_s$ stays out of equilibrium. Now, if the time between two weak interactions is large enough for active-sterile oscillations to take place, a non-trivial $\nu_s$ component develops due to oscillations. The next weak interaction collapses the neutrino state into another weak eigenstate, and leaves the sterile component unaffected. This continues on, and a non-zero abundance of sterile neutrinos build up. 

The time evolution of the $\nu_s$ phase-space distribution function $f_{\nu_s}(x, z)$ for a fixed energy $E\equiv x\,T$, where $T$ is the active neutrino temperature, is given by~\cite{Dodelson:1993je}
\begin{eqnarray}
\label{eq:masterequation}
\frac{d f_{\nu_s}}{d z} & = & \frac{\Gamma}{4H z} \, \frac{\Delta^2 \sin^22\theta}{\Delta^2 \sin^22\theta + \Gamma^2/4 + (\Delta \cos2\theta - V_T)^2}\, f_{\nu_a} \ .
\end{eqnarray}
Here, $\Gamma$ is a measure of the net $\nu_a$ interaction rate, $V_T$ quantifies the net thermal potential experienced by $\nu_a$ due to forward scattering in the plasma, and $\Delta \simeq m_4^2/(2E)$ is the vacuum oscillation frequency. Furthermore, we use the dimensionless variable $z\equiv\,1\,{\rm MeV}/T$ to have a measure of the time evolved. The $\nu_a$ distribution function, $f_{\nu_a}$, is taken to be Fermi-Dirac. The Hubble parameter is given by $H$. The DM relic is calculated by numerically integrating Eq.\,\ref{eq:masterequation} up to a value of $z$ where the $f_{\nu_s}$ contributes non-trivially.

However, this mechanism of producing DM is in tension with various observations, primarily from non-observation of x-rays by telescopes for $m_4 \gtrsim $~ few keV~\cite{Ng:2019gch}. On the other hand, sterile neutrino mass $m_4 \lesssim 2$~keV is inconsistent with phase-space bounds arising from dwarf-galaxies~\cite{Tremaine:1979we}. This is where non-standard neutrino self-interactions come to the rescue!

\section{Non-standard neutrino self-interactions}
We add to the SM the following interaction, $\mathcal{L}	\supset\frac{\lambda_\phi}{2}\, \nu_a \nu_a \phi + {\rm h.c.} $,
where $\phi$ is a complex scalar, having a mass $m_{\phi}$. For this work, we stay agnostic of how this interaction can be embedded in an ultraviolet-complete models. This new interaction contributes additional production channels for sterile neutrinos by introducing additional terms in $\Gamma$ and $V_T$. The details of the computation are outlined in~\cite{DeGouvea:2019wpf}. The mechanism works like a seesaw: if the rate of new interactions producing $\nu_a$ is increased, the same quantity of $\nu_s$ can be produced with a smaller mixing angle, and hence the astrophysical bounds can be evaded.
%%%
%%%
\subsection{$m_\phi>\,{\rm few\,\, MeV}$}
%%%
%%%
In this case, there can be two possible contributions to $\Gamma$: in the limit $m_\phi \gg T$, we have $\Gamma\propto T^5/m_\phi^4$, while a lighter mass $\phi$ ($T\gtrsim m_\phi$) can be produced in the plasma with $\Gamma \propto T$. In all these cases, the total rate is the sum of these new interaction rates and the usual SM interaction rate. Similarly, for the thermal potential, along with the SM terms, there are additional contributions
$V_T \propto  (T^5/ m_\phi^4)$ for $m_\phi \gg T$, and
$V_T \propto T$ for $m_\phi \ll T$.

\begin{figure}[!t]
\includegraphics[height=0.33\linewidth,width=0.44\linewidth]{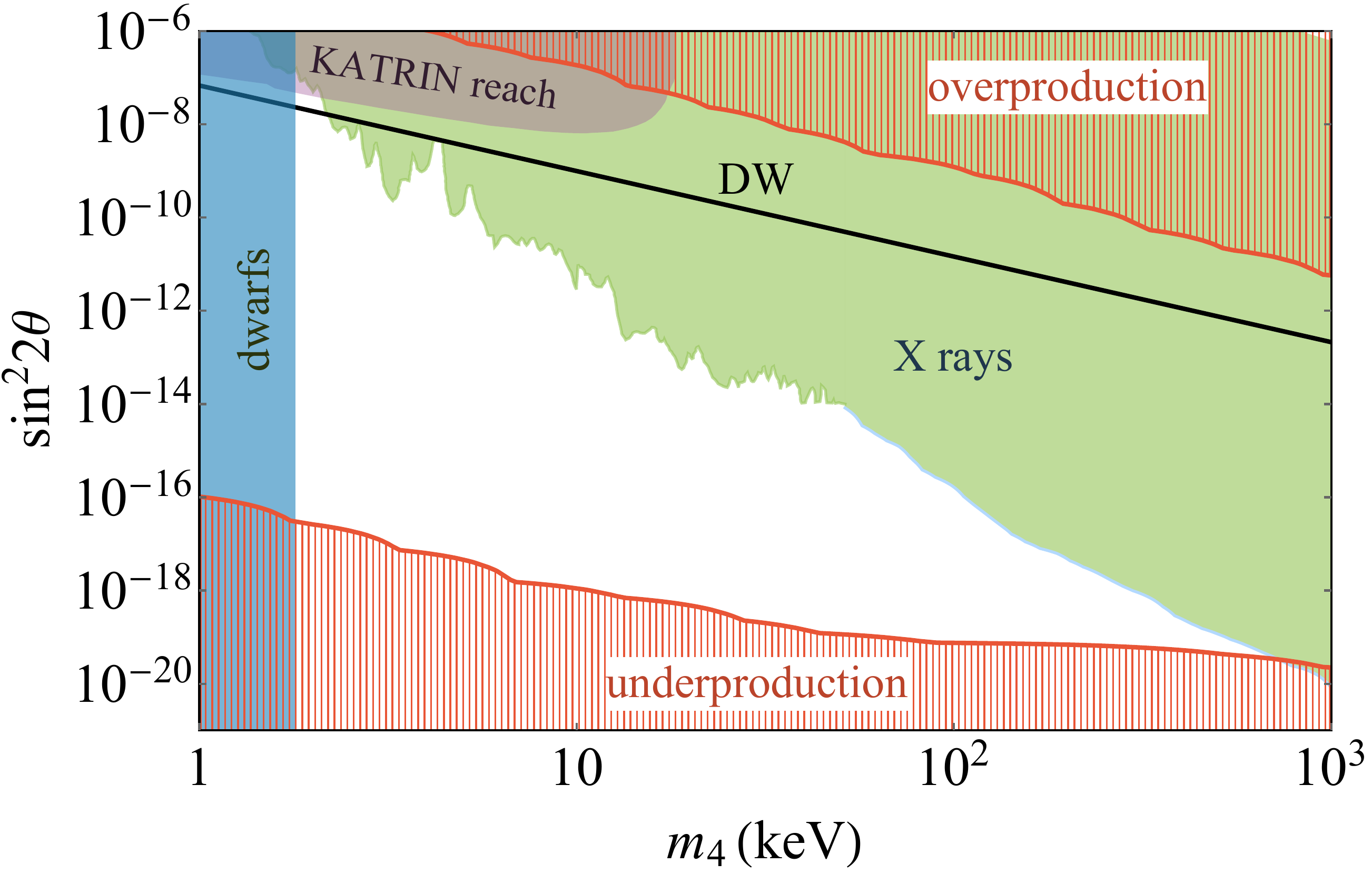}~~\includegraphics[height=0.345\linewidth,width=0.44\linewidth]{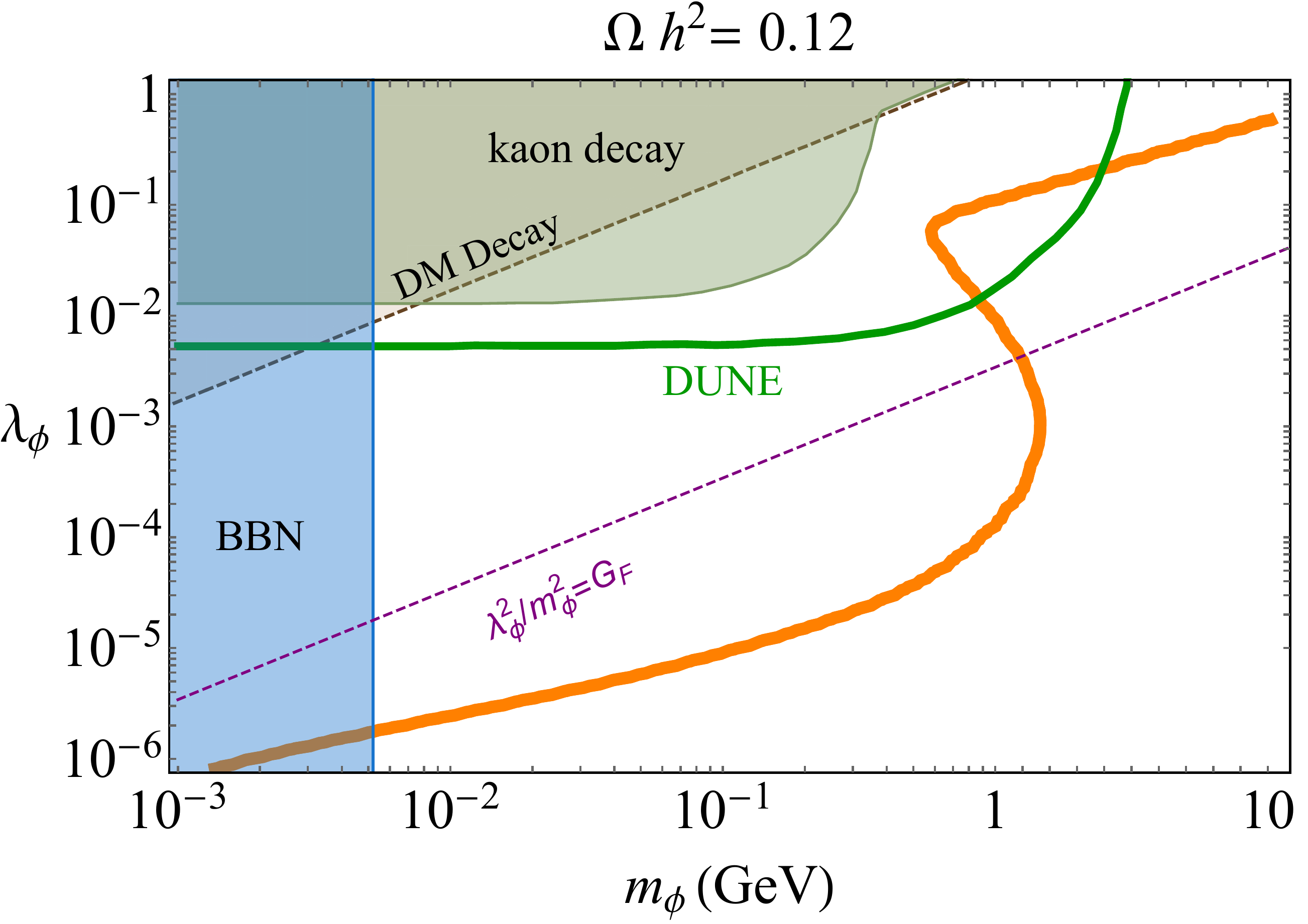}
	\caption{Left: Region of the $m_4$- $\sin^2 2\theta$ parameter space where one can find values of $(m_\phi, \lambda_{\phi})$ such that the sterile neutrino accounts for all the DM. Constraints from X-ray searches, KATRIN AND dwarf galaxies are also shown. The hatched red regions do not produce the correct relic for any choice of $(m_\phi, \lambda_{\phi})$. Right: Parameter space of $m_\phi-\lambda_\phi$ for $m_4=7.1\,{\rm keV}$ and $\sin^22\theta=7\times10^{-11}$, where new neutrino self-interactions can make up all of DM (orange line). The shaded regions show constraints from meson decays (green), BBN (blue), and regions where the DM decays too soon (dashed). }
	\label{SterileDMSearch}
\end{figure}

Using these ingredients, we integrate Eq.\,\ref{eq:masterequation} up to $z\sim 10$, where $\nu_s$ are still relativistic to compute the DM relic density. Fig.~\ref{SterileDMSearch} (left) shows the region of the $m_4-\sin^2 2\theta$ parameter space for a range of $(m_\phi,\lambda_{\phi})$ values, such that the $\nu_s$ accounts for all the DM. The bounds coming from x-rays are shown in green\cite{Ng:2019gch}, those from dwarf galaxies in blue~\cite{Tremaine:1979we}, while the predicted reach of KATRIN in magenta~\cite{Mertens:2015ila}. The black line (DW) corresponds to the region predicted by the DW mechanism to satisfy the relic density; clearly this is in tension with the current bounds. However, in presence of extra neutrino self-interactions, the parameter space between the two hatched red regions opens up. Thus, for any point in the white region, there exists a collection of $( m_\phi,\lambda_\phi)$ values that satisfy the DM relic. 

Secret neutrino self-interactions stronger than the weak interactions can be probed in laboratory experiments, as shown in Fig.~\ref{SterileDMSearch} (right). The orange ``S"-shaped curve shows the region in the $m_\phi-\lambda_\phi$ parameter space, where the correct DM relic density is satisfied. Meson decays provide stringent constraints, due to the presence of three neutrino final states, $M^+\to l^+ \bar{\nu}_{l} + (\phi \to \nu\nu)$. For example, if the neutrino is a $\nu_\mu$, the constraints coming from Kaon decays are shown in green~\cite{E949:2016gnh}. This also dominates over constraints coming from invisible decays of $Z$, and Higgs. For $m_\phi$ below a few MeV, big-bang nucleosynthesis (BBN) predictions are altered due to enhanced $\phi$ production (blue). Finally, the upcoming Deep Underground Neutrino Experiment (DUNE) will be very sensitive to these neutrino self-interactions, through the radiation of a light $\phi$ in neutrino matter interactions (solid green line)~\cite{Berryman:2018ogk}. 
%%%
%%%
\subsection{$m_\phi< \,{\rm few\,\, MeV}$}
%%%
%%%
In this case, the mediator has to couple to neutrinos feebly to avoid bounds arising from BBN. We focus on a scenario where the coupling is weak enough such that $\phi$ enters into thermal equilibrium with $\nu_a$ after BBN. The $\nu_s$-DM is produced primarily through new neutrino self-interactions mediated by an on-shell $\phi$. To avoid bounds arising from measurement of extra radiation during the cosmic microwave background (CMB) epoch through the quantity $\Delta N_{\rm eff}$, these scalars must decay before CMB. Post-BBN thermalization of $\phi$ ends up modifying standard cosmology, thereby altering the predictions for  $\Delta N_{\rm eff}$ both during BBN and CMB. Hence, the allowed parameter space for $\nu_s$-DM relic density has a strong correlation with $\Delta N_{\rm eff}$.

The thermalization of $\phi$ after the BBN epoch is calculated by solving the following Boltzmann equation, 
\begin{equation}\label{eq:PhiFreezeIn}
\frac{d Y_\phi}{dz} = - n_\nu^{eq}\frac{\langle (m_\phi/E) \Gamma_{\phi\rightarrow \nu \nu} \rangle}{s H z} \left( \frac{Y_\phi}{Y_\phi^\text{\sc eq}}  - 1 \right) \ ,
\end{equation}
where $Y_\phi=n_\phi/s$ is the $\phi$ yield with $n_\phi$ being the number density, and $s$ the entropy density of the Universe. Here $\Gamma_\phi = 3\lambda^2 m_\phi/(32\pi)$ is the $\phi$ decay rate at rest, and $n_\nu^{eq}$ is the number density of active neutrinos in equilibrium. As the $\phi-\nu_a$ system reaches equilibrium, a chemical potential develops for $\phi$ and $\nu_a$. Furthermore, the neutrino temperature also deviates from its value computed from standard cosmology. Both these quantities can be computed using energy/lepton-number conservation~\cite{Kelly:2020aks}. Since $\phi$-equilibrium is irreversible, entropy is not conserved in the process. During this period, $\nu_s$-DM is produced due to on-shell scattering $\nu\nu\to\nu \nu_4$ mediated by $\phi$, until a time comes when $\phi$ becomes non-relativistic and freezes out.

\begin{figure}[!t]
  \includegraphics[height=0.285\linewidth,width=0.44\linewidth]{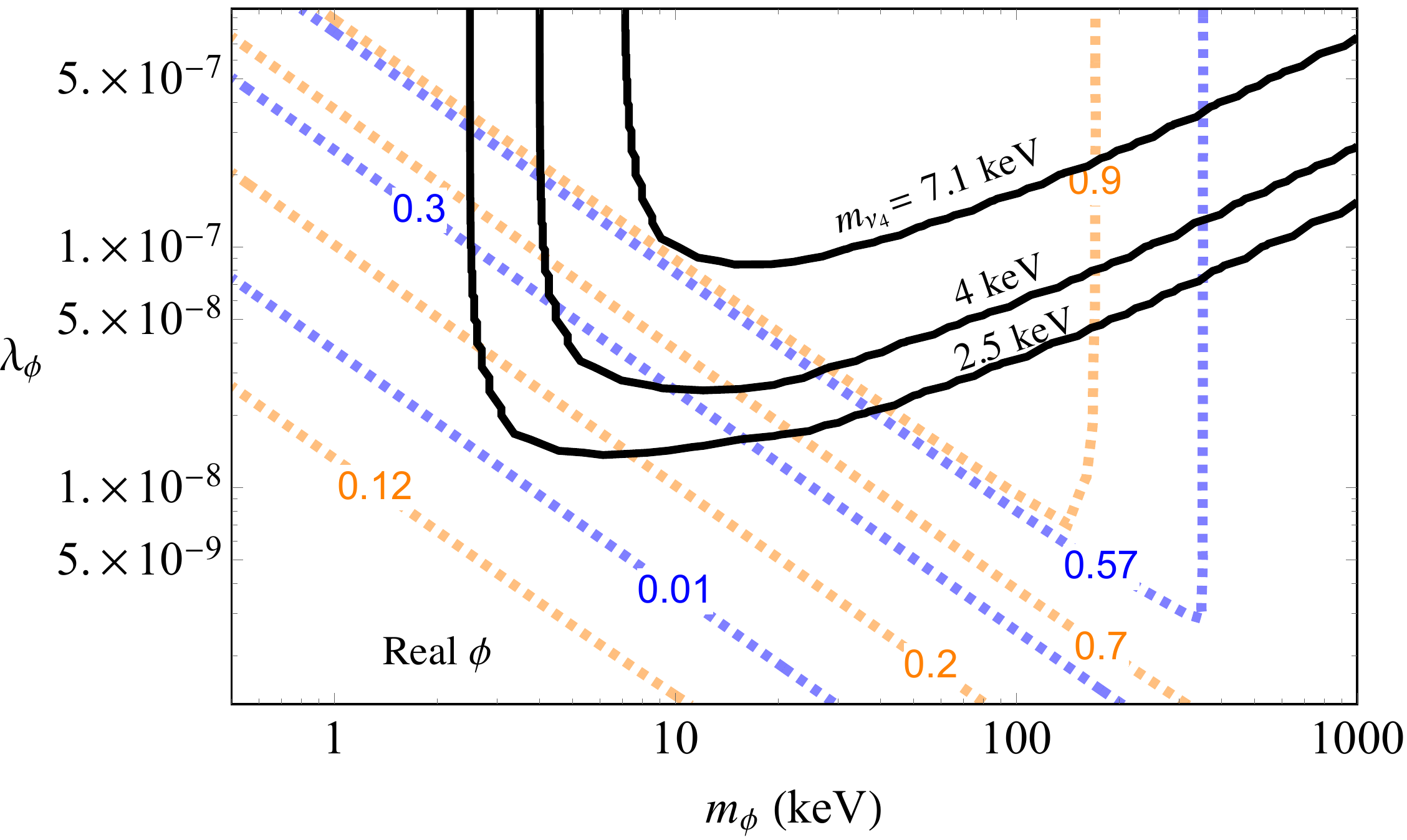}~~\includegraphics[height=0.29\linewidth,width=0.44\linewidth]{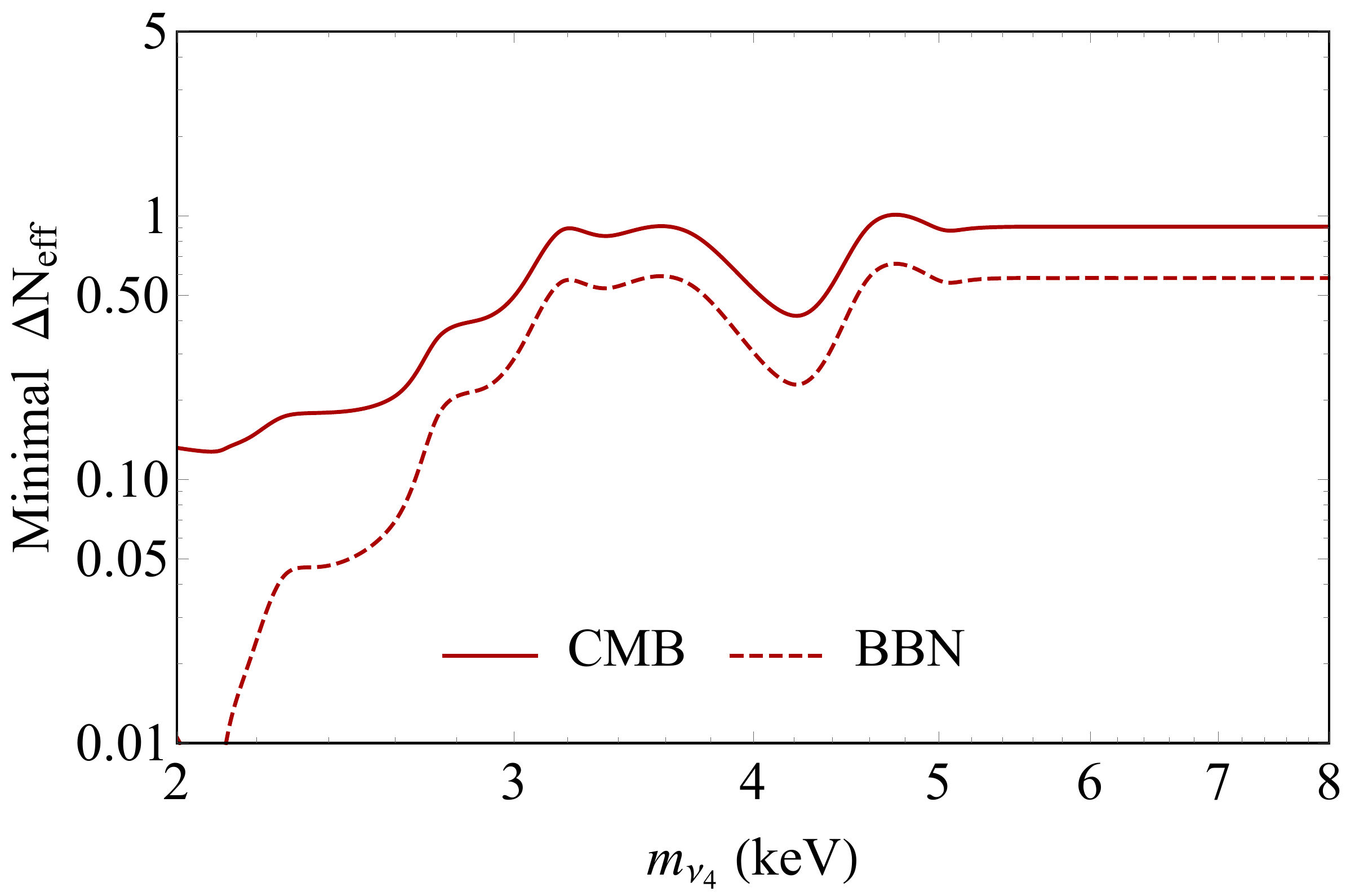}
    \caption{Left: Contours satisfying the correct $\nu_s$DM relic density (black) for different values of $m_4$, and $\Delta N_{\rm eff}$ contours for CMB and BBN (orange and blue) in the $m_\phi-\lambda_\phi$ plane. Right: Minimum value of $\Delta N_{\rm eff}$ at CMB (solid) and BBN (dashed) for different values of $m_4$. These contours correspond to the largest value of $\theta$ allowed by current x-ray searches. 
}
    \label{fig:LambdaMPhi}
\end{figure}

Using this formalism, one can calculate the relic density of $\nu_s$-DM using Eq.\,\ref{eq:masterequation} in the limit of small $m_\phi$ and $\lambda_\phi$, where the effective active-sterile mixing angle is almost the same as the vacuum one. We show the results in Fig.~\ref{fig:LambdaMPhi} (left) for three values of the DM masses, $m_4=\left\lbrace2.5,\, 4,\, 7.1\right\rbrace\,$keV, which satisfy the relic density (black lines). For $m_\phi \gg m_4$,  the lines show a $\lambda \sim \sqrt{m_\phi}$ behaviour, in tandem with the results for $m_\phi\gtrsim$\,few MeV. As $m_\phi \to m_4$, phase-space suppression forces the curves to bend up. Constant contours of $\Delta N_{\rm eff}^\text{\sc bbn}$ and $\Delta N_{\rm eff}^\text{\sc cmb}$ generated by the new interaction are also shown. For each value of $\nu_s$-DM mass, it is possible to derive a lower limit on the values of $\Delta N_{\rm eff}^\text{\sc bbn, cmb}$. As a result, for a given DM mass, one can combine the current x-ray bounds on the mixing angle, and derive the minimum values of  $\Delta N_{\rm eff}^\text{\sc bbn,cmb}$ which can be generated. This is shown in Fig.~\ref{fig:LambdaMPhi} (right), where we plot the minimal values of $\Delta N_{\rm eff}^\text{\sc bbn,cmb}$ which can be probed for a range of $m_4$ values. We find that  $\Delta N_{\rm eff}^\text{\sc cmb}>0.12$, which is definitely a target for the upcoming CMB-S4 experiment~\cite{CMB-S4:2016ple}. Negligible production of $\phi$ during BBN does not allow such an absolute lower limit during BBN. Such an interplay between constraints from upcoming x-ray telescopes and cosmology probes will play an exciting role in the discovery of sterile neutrino dark matter. 
\begin{figure}[!t]
\centering
\includegraphics[width=0.48\textwidth]{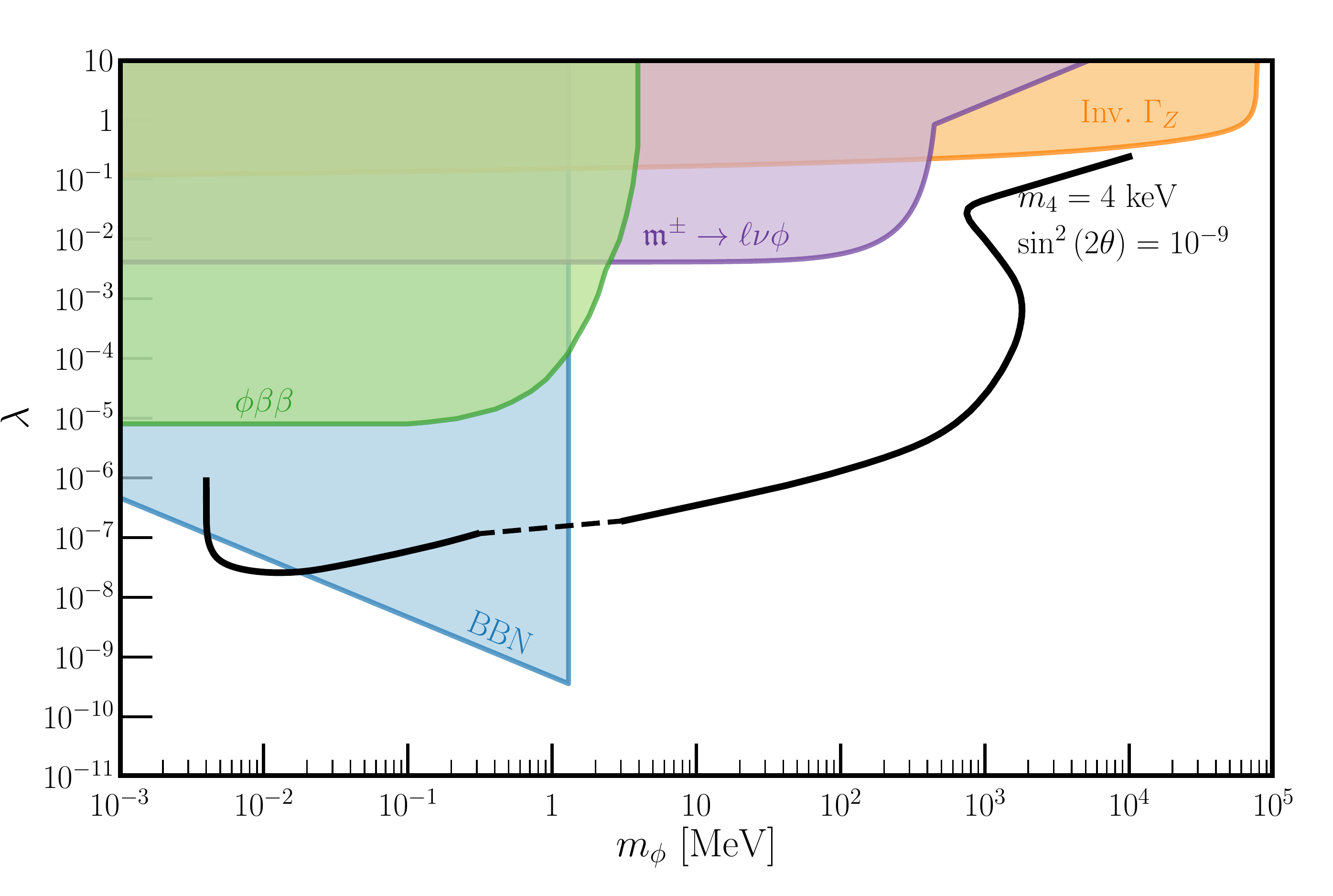}
	\caption{The Big Picture, showing the full parameter space for $\nu_s$DM that can account for the relic density with the corresponding constraints.}
	\label{BigPicture}
\end{figure} 
\section{The Big Picture}
To summarize, we studied different ways of efficiently producing sterile neutrino DM in the early Universe, through active neutrino self-interaction, mediated by a scalar $\phi$. These neutrino self-interactions, if stronger  than the SM weak interactions, are testable, and can leave an impact on different laboratory probes, including the upcoming DUNE experiment. On the other hand, feeble self-interactions can even allow a sub-GeV scalar to mediate self-interactions. These models can be probed in upcoming cosmology surveys like CMB-S4 due to the amount of extra radiation produced due to late thermalization and decay of the scalar. These two results can be nicely combined into a big picture, as shown in Fig.~\ref{BigPicture}.

The black solid curve shows the entire parameter space of the theory, where the DM relic density can be satisfied for a light sterile neutrino of mass 4 keV. The history of evolution of the $\nu_s$-DM in the early Universe is different in the region above and below MeV scale. This accounts for the dashed joining line in the region around the MeV scale. The strongest probes in the region where the mediator is above MeV scale come from terrestrial experiments (meson decays, Z decays). On the other hand, for sub-MeV mediators, the best probes are from cosmological observations of $\Delta N_{\rm eff}$ and neutrinoless double beta decays. Combined, this serves as an excellent target for upcoming experiments, from terrestrial laboratories to cosmological surveys.

\section{Acknowledgments}
M.S. wishes to thank his collaborators Andre de Gouvea, Kevin Kelly, Walter Tangarife and Yue Zhang for valuable contributions in the original works~\cite{DeGouvea:2019wpf,Kelly:2020pcy,Kelly:2020aks}. M.S. acknowledges support from the National Science Foundation, Grant PHY-1630782, and to the Heising-Simons Foundation, Grant 2017-228, during the completion of these works. M.S. also thanks the MPIK, Heidelberg for hospitality during the writing of this manuscript.

%%%%%%%%%%%%%%%%%%%%%%%%%%%%%%%%%%%%%%%%%%%

\end{document}